# New Supersymmetrizations of the Generalized KdV Hierarchies

José M. Figueroa-O'Farrill [†] [‡]

and

Sonia Stanciu[§]

*Physikalisches Institut der Universität Bonn*
*Nußallee 12, W–5300 Bonn 1, GERMANY*

## ABSTRACT

Recently we investigated a new supersymmetrization procedure for the KdV hierarchy inspired in some recent work on supersymmetric matrix models. We extend this procedure here for the generalized KdV hierarchies. The resulting supersymmetric hierarchies are generically nonlocal, except for the case of Boussinesque which we treat in detail. The resulting supersymmetric hierarchy is integrable and bihamiltonian and contains the Boussinesque hierarchy as a subhierarchy. In a particular realization, we extend it by defining supersymmetric odd flows. We end with some comments on a slight modification of this supersymmetrization which yields local equations for any generalized KdV hierarchy.

---

[†]  e-mail: jmf@avzw03.physik.uni-bonn.de

[‡]  Address after September 1993: Physics Department, Queen Mary and Westfield College, London, UK.

[§]  e-mail: sonia@avzw03.physik.uni-bonn.de

## Introduction

Recently, a new supersymmetric extension of the KdV hierarchy has appeared in the context a matrix-model-inspired approach to $2d$ quantum supergravity [1]. We recognized in [2] that this hierarchy is but the KdV hierarchy in disguise—the KdV variable being replaced by an even superfield. This then allowed us to write the flows in a manifestly supersymmetric fashion, to find a bihamiltonian structure, and to construct an infinite number of conserved supersymmetric charges in involution—thus formally proving integrability.

This result raises the question whether this supersymmetrization works for the generalized KdV hierarchies. This question is interesting in view of its applications to non-critical superstrings, as well as from the the general theory of supersymmetric integrable systems. We will show in this Letter that the supersymmetrization in [2] works only in the case of the Boussinesque hierarchy, whereas a different and—in a sense—more natural supersymmetrization works for all cases. These more general supersymmetric hierarchies are in a sense not new, since one can prove that they are particular reductions of the known supersymmetric KP hierarchies [3] [4] [5]. Nevertheless their bihamiltonian structures do not arise in this way, and the conserved charges are constructed in a novel fashion that has features which hold some interest in their own right.

To understand the idea behind these new supersymmetrizations, let us briefly recall the main features of the generalized KdV hierarchies. The $n$-KdV hierarchy is defined as the isospectral flows of the differential operator $L = \partial^n + \sum_{i\geq 2} u_i \partial^{n-i}$. The flows are given by equations of the form

$$\frac{\partial u_i}{\partial t_j} = P_{ij}(u) \tag{1}$$

where the $P_{ij}$ are differential polynomials in the $\{u_i\}$. These flows are then extended as evolutionary derivations—*i.e.*, derivations commuting with $\partial$—to the whole differential ring $R[u_i]$ generated by the $\{u_i\}$. Therefore, formally, the $n$-KdV hierarchy is defined on any differential ring which is freely generated by abstract variables $\{u_i\}$. One can go a long way along this formal path. First of all, one can prove that the flows commute. Furthermore, using the formal calculus of variations, one can then define bihamiltonian structures, construct conserved charges in involution, and prove the formal integrability of the hierarchies (see, for example, [6]). It is only when discussing solutions of the evolution equations that one is forced to choose a concrete realization for the differential ring $R[u_i]$ as a subring, say, of the (rapidly decaying, periodic,...) smooth functions on the real line.

As discussed in [2], the supersymmetric extension of the KdV hierarchy ($n = 2$) discovered in [1]—hereafter referred to as SKdV-B—is obtained by replacing the KdV variable $u$ by the even superfield[1] $U' = u + \theta\tau'$. Since $U'$ freely generates a differential ring, we are well within the domain of the formal KdV hierarchy and, in particular, this means that all the above mentioned results carry over.

---

[1] We use the following conventions on derivatives: acting on superfields, a prime denotes the action of the supercovariant derivative $D = \frac{\partial}{\partial\theta} + \theta\partial$; whereas otherwise it simply denotes the action of $\partial$. This should not cause any confusion.

There are two caveats, however. First of all, we want to interpret these flows as those from a supersymmetric hierarchy. This means that we cannot simply take the conserved charges to be the ones that would follow from the KdV hierarchy with $U'$ replacing $u$, since these still have $\theta$-dependence. In fact, each KdV conserved charge furnish us now with two conserved charges, since both the $\theta$-dependent and the $\theta$-independent parts are separately conserved. Only one of them, however, is invariant under supersymmetry and is the one that we would understand as the supersymmetric conserved charge. The second caveat is that since it is $\tau'$ that enters in the superfield, the evolution equations will be equations for $\tau'$. We must then make sure that the resulting equations for $\tau$ are actually local.

As we showed in [2] neither of these two problems prevent the supersymmetrization of the KdV hierarchy, and we will show in this Letter that neither are the analogous problems present for the supersymmetrization of the Boussinesque ($n$=3) hierarchy. For $n > 3$, however, the resulting equations for the superpartners of the $u_i$ are in general nonlocal and we are forced to conclude that the supersymmetrization does not work.

One may wonder why it is that we replace the $u_i$ by $U'_i$ and not simply by even superfields $V_i = u_i + \theta \sigma_i$. From the point of view of supersymmetric integrable systems, of course, there is no reason not to consider these hierarchies which, in fact, appear much more natural. However, they do not seem to be the ones that are interesting in view of their applications to superstring theory. Nevertheless, for completeness, we devote a short section to this more natural supersymmetrization.

This paper is thus organized as follows. We first define the supersymmetric extension SBs of the Boussinesque hierarchy and then we prove its formal integrability; that is, we construct an infinite number of independent nontrivial conserved charges and a bihamiltonian structure relative to which the conserved charges are in involution. We then define a particular realization of the SBs hierarchy similar to the realization of SKdV-B appearing in the supersymmetric analog of the one-matrix model. This allows us to extend the hierarchy by defining odd supersymmetric flows as well as to anticipate a realization of this hierarchy in the (still to be constructed) supersymmetric analog of the two-matrix model. We then prove that the supersymmetrization of the ($n > 3$)-KdV hierarchies result in nonlocal equations. And finally, we conclude this Letter with some brief comments on the more natural supersymmetrization alluded to above.

The SBs Hierarchy

Consider the Lax operator $L = \partial^3 + u\partial + v$ and let $R[u,v]$ denote the differential ring freely generated by its coefficients. The isospectral flows of $L$ define on $R[u,v]$ the Boussinesque hierarchy. These flows can be written in a way that exhibits the Lenard recursion relations for the conserved charges:

$$\begin{pmatrix} \frac{\partial u}{\partial t_r} \\ \frac{\partial v}{\partial t_r} \end{pmatrix} = J_1 \cdot \begin{pmatrix} \frac{\delta H_{r+3}}{\delta u} \\ \frac{\delta H_{r+3}}{\delta v} \end{pmatrix} = J_2 \cdot \begin{pmatrix} \frac{\delta H_r}{\delta u} \\ \frac{\delta H_r}{\delta v} \end{pmatrix} , \qquad (2)$$

where $H_r \equiv \frac{3}{r} \operatorname{Tr} L^{3/r}$, $r \in \mathbf{N}$, $r \not\equiv 0 \mod 3$ are the conserved charges, and the $J_i$'s are the two coordinated hamiltonian structures, relative to which the $H_r$'s are in involution:

$$J_1 = \begin{pmatrix} 0 & 3\partial \\ 3\partial & 0 \end{pmatrix} , \qquad (3)$$

and

$$J_2 = \begin{pmatrix} -\frac{2}{3}(\partial^5 + \partial^3 u + u\partial^3 + u\partial u) + \partial^2 v - v\partial^2 & \partial^4 + u\partial^2 + \partial v + 2v\partial \\ -\partial^4 - \partial^2 u + 2\partial v + v\partial & 2\partial^3 + \partial u + u\partial \end{pmatrix} . \qquad (4)$$

Following our discussion above, the Boussinesque hierarchy is defined abstractly in $R[u,v]$ and to obtain a concrete realization of the hierarchy we must choose a representation of this differential ring. Let us consider the particular realization of the abstract Boussinesque hierarchy where the generators $u$ and $v$ of the differential ring are taken to be even superfields $U'$ and $V'$, respectively. We shall denote this hierarchy by $\mathsf{Bous}(U', V')$. We define then our supersymmetric extension (SBs) of the Boussinesque hierarchy as the hierarchy induced by $\mathsf{Bous}(U', V')$ on the superdifferential ring $S[U,V]$. More concretely, the flows of this new hierarchy are obtained from the flows of $\mathsf{Bous}(U', V')$ by stripping off one derivative from both sides of each equation. A natural question now appears: are these flows local? Fortunately the answer is positive, which can be easily shown if we consider the first hamiltonian structure of the Boussinesque hierarchy (3). Then the new flows read

$$\begin{pmatrix} \frac{\partial U}{\partial t_r} \\ \frac{\partial V}{\partial t_r} \end{pmatrix} = D^{-1} J_1 \cdot \begin{pmatrix} \left.\frac{\delta H_{r+3}}{\delta u}\right|_{\substack{u=U' \\ v=V'}} \\ \left.\frac{\delta H_{r+3}}{\delta v}\right|_{\substack{u=U' \\ v=V'}} \end{pmatrix} = \begin{pmatrix} 3D \cdot \left.\frac{\delta H_{r+3}}{\delta u}\right|_{\substack{u=U' \\ v=V'}} \\ 3D \cdot \left.\frac{\delta H_{r+3}}{\delta v}\right|_{\substack{u=U' \\ v=V'}} \end{pmatrix} . \qquad (5)$$

Let us introduce the respective superpartners $\tau$ and $\sigma$ of $u$ and $v$, by

$$U = \tau + \theta u \quad \text{and} \quad V = \sigma + \theta v . \qquad (6)$$

These fields carry the standard representation of the supersymmetry algebra:

$$\delta u = \tau' \qquad \delta v = \sigma' \qquad \delta \tau = u \qquad \delta \sigma = v , \qquad (7)$$

which in terms of superfields will be given by

$$\delta U = Q \cdot U \quad \text{and} \quad \delta V = Q \cdot V , \qquad (8)$$

where $Q \equiv \frac{\partial}{\partial \theta} - \theta \partial$. Writing the flows for $U$ and $V$ in components, we of course recover the original Boussinesque hierarchy for the even fields $u$ and $v$. This suggests an alternative description of this supersymmetrization procedure.

We first consider the embedding $R[u,v] \to S[U,V]$ given by

$$u \mapsto U' = u + \theta \delta u \quad \text{and} \quad v \mapsto V' = v + \theta \delta v , \tag{9}$$

so that the variables of $\mathsf{Bous}(U',V')$ can be thought of as some sort of deformation of the Boussinessque variables. And then we introduce the odd fields $\tau, \sigma$ such that they carry the above mentioned representation of the supersymmetry algebra. Hence the odd superfields generating $S[U,V]$ will be those given by (6) and the supersymmetric hierarchy in which we are interested is the one induced on them.

Conserved Charges, Bihamiltonian Structure, and Integrability

First we need a technical lemma concerning the variational derivatives before and after the embedding $R[u,v] \to S[U,V]$.

LEMMA 10. *Let $h(u,v) \in R[u,v]$. Then, in $S[U,V]$,*

$$\frac{\delta}{\delta U} \int_B h(U',V') = D \cdot \left.\frac{\delta}{\delta u} \int h(u,v)\right|_{\substack{u=U' \\ v=V'}} ,$$

*and the analogous relation for $V$.*

PROOF: By *definition*, for any $h \in S[U,V]$,

$$\frac{\delta}{\delta U} \int_B h = \sum_{i \geq 0} (-)^{i(i+3)/2} D^i \cdot \frac{\partial h}{\partial U^{[i]}} , \tag{11}$$

whence if $h$ only depends on the $U^{[\text{odd}]}$,

$$\frac{\delta}{\delta U} \int_B h = \sum_{k \geq 0} (-)^k D^{2k+1} \frac{\partial h}{\partial U^{[2k+1]}}$$

$$= D \cdot \sum_{k \geq 0} (-)^k \partial^k \frac{\partial h}{\partial (U')^{(k)}} ,$$

which proves the lemma. ∎

The next proposition tells us how the conserved charges of the SBs hierarchy come induced from those of Boussinesque.

PROPOSITION 12. *If $H_{3k+r} \equiv \frac{3}{r} \operatorname{Tr} L^{k+r/3}$, $k = 0, 1, 2, ...$, $r = 1, 2$ are the conserved charges of the Boussinesque hierarchy, then the corresponding conserved charges for the SBs hierarchy are given by*

$$K_{3k+r} = \delta H_{3k+r}, \qquad \text{if } k = 1, 2, \ldots , \tag{13}$$

*whereas*

$$K_r = H_r, \qquad \text{for } r = 1, 2 \tag{14}$$

*and for $k \geq 1$ they obey the following relation:*

$$\tfrac{3}{r} \operatorname{Tr} L^{k+r/3} = H_{3k+r}[u,v] + \theta K_{3k+r}[u,v,\tau,\sigma] . \tag{15}$$

*Moreover, the conserved charges are independent.*

PROOF: For the case $K_{3k+r} \neq 0$ we send the reader to [**2**] where this point, although discussed in the context of the KdV hierarchy, has been proven for any generalized $n$-KdV hierarchy. Hence the only thing that remains to be shown is that these charges are indeed nontrivial. Suppose then that $K_{3k+r} = 0$. Then using Lemma 10 we also have that

$$\frac{\delta}{\delta U} \int_B h_{3k+r}(U,V) = D \cdot \frac{\delta}{\delta u} \int h_{3k+r} \bigg|_{u=U'} = 0 \qquad (16)$$

and the analogous relation for $v$. But since the conserved charges of the Boussinesque hierarchy are nontrivial it follows necessarily that $\frac{\delta}{\delta u} H_{3k+r}$ is a constant. Knowing that $H_{3k+r}$ has weight $3k+r$ (where we say that $\partial$ has weight 1, and $u$ and $v$ have weights 2 and 3, respectively) it is clear that the only case in which we can get a constant is when $k=0$. Noticing that $H_1 \propto \int u$ and $H_2 \propto \int v$, and that $\int_B U$ and $\int_B V$ are actually conserved, we can then conclude that the first two conserved charges are Berezinians of the two generating superfields themselves, rather than of their first derivatives. Finally, and using the grading just introduced, it is easy to see that all the charges have different degrees, and hence are independent. ∎

In order to prove the integrability of the SBs hierarchy, we first determine its bihamiltonian structure, which again comes induced from the bihamiltonian structure of the Boussinesque hierarchy.

PROPOSITION 17. *The SBs hierarchy inherits a bihamiltonian structure from that of the Boussinesque hierarchy.*

PROOF: Considering the bihamiltonian structure (3) and (4) of the Boussinesque hierarchy, substituting $u$ and $v$ for $U'$ and $V'$, stripping off the derivatives, and using Lemma 10, we obtain

$$\begin{pmatrix} \frac{\partial U}{\partial t_r} \\ \frac{\partial V}{\partial t_r} \end{pmatrix} = J_1^S \cdot \begin{pmatrix} \frac{\delta H_{r+3}}{\delta U} \\ \frac{\delta H_{r+3}}{\delta V} \end{pmatrix} = J_2^S \cdot \begin{pmatrix} \frac{\delta H_r}{\delta U} \\ \frac{\delta H_r}{\delta V} \end{pmatrix}, \qquad (18)$$

where $J_i^S = D^{-1} \cdot J_i \cdot D^{-1}$. Notice that this is already suggests a bihamiltonian structure. Indeed, notice that $J_1^S$ satisfies the Jacobi identities trivially, since it is constant. The second structure $J_2$ may not seem obviously Poisson, but it is not hard to show that the Jacobi identities are also satisfied; although it defines a nonlocal bracket. Finally, notice that $J_1$ can be obtained from $J_2$ by shifting $V' \mapsto V' + \lambda$. Since $J_2$ is Poisson for any $U$ and $V$, it follows that $J_1$ and $J_2$ are coordinated. ∎

Usual arguments now imply that the conserved charges are in involution relative to both Poisson structures. In summary, SBs is a (formally) integrable bihamiltonian supersymmetric hierarchy.

Odd flows and the Supersymmetric Analog of the Two-Matrix Model

So far we managed to define an integrable supersymmetric extension of the Boussinesque hierarchy that consists only of even flows on the superdifferential ring $S[U,V]$. A natural question to ask is whether one can actually extend this hierarchy in a supersymmetric fashion by odd flows. This point, apart from being important for the general theory of integrable systems, is of particular interest in the context of the matrix model approach to (noncritical) superstrings.

It is the purpose of this section to show that, in a particular realization, one can actually define supersymmetric odd flows for the SBs hierarchy. This realization is similar to the realization of the SKdV-B hierarchy appearing in the supersymmetric analog of the one-matrix model [1] and it is to be expected that it is the realization that will appear in the supersymmetric analog of the two-matrix model, if one exists.

The realization in question is defined as follows. Let us introduce an infinite number of odd "times" $\tau_n$ and let the dependence of $u$, $v$, $\tau$ and $\sigma$ on these times be given by

$$\begin{array}{ll} \frac{\partial u}{\partial \tau_n} = 0 \ , & \frac{\partial \tau}{\partial \tau_n} = A_n(u,v) \ , \\ \frac{\partial v}{\partial \tau_n} = 0 \ , & \frac{\partial \sigma}{\partial \tau_n} = B_n(u,v) \ , \end{array} \tag{19}$$

where $A_n(u,v)$ and $B_n(u,v)$ are arbitrary even functions to be fixed presently. One can then formally integrate the equations for the odd fields writing them, in this way, as a linear combination of the odd times:

$$\tau = \sum_k \tau_k A_k(u,v) \qquad \text{and} \qquad \sigma = \sum_k \tau_k B_k(u,v) \ . \tag{20}$$

The functions $A_n$ and $B_n$ are uniquely defined if we demand that the odd and even times are related by supersymmetry in the following way:

$$[\delta, \frac{\partial}{\partial t_n}] = 0 \qquad \text{and} \qquad [\delta, \frac{\partial}{\partial \tau_n}] = -\frac{\partial}{\partial t_{n-1}} \ . \tag{21}$$

Indeed, applying the above formal relation to $u$ and $v$ and using the form of the flows of the Boussinesque hierarchy one obtains an explicit expression for the odd fields in terms of the even ones

$$\tau = -3 \sum_k \tau_k \frac{\delta H_{k+1}}{\delta v} \ , \qquad \sigma = -3 \sum_k \tau_k \frac{\delta H_{k+1}}{\delta u} \ . \tag{22}$$

This is not all, though, since $\tau$ and $\sigma$ still have to satisfy the supersymmetry transformation laws (7). This imposes conditions on the transformation properties of the $\tau_k$ under supersymmetry, which can be satisfied (see [2]) if $\delta \tau_k = -\delta_{k,1}$. It is then clear that the following (odd) flows

$$D_n \equiv \frac{\partial}{\partial \tau_n} - \tau_1 \frac{\partial}{\partial t_{n-1}} \tag{23}$$

are supersymmetric and commute with the even flows; whereas among themselves they obey the following algebra:

$$D_1^2 = -\partial \quad \text{and} \quad [D_1, D_n] = -\frac{\partial}{\partial t_{n-1}}, \quad \text{for } m \geq 2 . \tag{24}$$

This defines a supersymmetric extension of the SBs hierarchy by odd flows. As mentioned in [**2**] in the context of the SKdV-B hierarchy, there are obstacles in writing the flows in superspace. This remains an interesting open problem.

Supersymmetrization of $(n > 3)$-KdV

We have worked out in detail the first two cases of the generalized KdV hierarchies and we have seen how the supersymmetrization procedure that we defined leads to integrable extensions whose structure is closely related to the one of the original hierarchies. This success is nevertheless tied to a rather remarkable feature of the first two hierarchies—namely, that the image of their first hamiltonian structure $J_1$ are perfect derivatives, making possible that the new flows be local.

In order to see that this feature breaks down for the higher hierarchies, let us first consider the 4-KdV hierarchy. This hierarchy is defined by the Lax operator $L = \partial^4 + u_2 \partial^2 + u_3 \partial + u_4$ and its first hamiltonian structure is given by

$$J_1 = \begin{pmatrix} 2\partial^3 + u_2 \partial + \partial u_2 & -2\partial^2 & 4\partial \\ 2\partial^2 & 4\partial & 0 \\ 4\partial & 0 & 0 \end{pmatrix} . \tag{25}$$

As one can already see, unlike for the KdV and Boussinesque hierarchies, here $J_1$ depends explicitly on the fields $u_i$ and acting on a generic gradient, will not yield a perfect derivative. Of course, this is not enough to conclude that the flows of the supersymmetric extension of this hierarchy are not local: there exist hierarchies of local equations with nonlocal hamiltonian structures (*cf.* SKdV). We must therefore look at the explicit form of the flows. If one writes down the second flow, for instance, one gets for $u_4 \mapsto U_4'$

$$\frac{\partial U_4'}{\partial t_2} = U_4^{[5]} - \tfrac{1}{2} U_2^{[9]} - \tfrac{1}{2} U_2' U_2^{[5]} - \tfrac{1}{2} U_2^{[3]} U_3' , \tag{26}$$

which induces a nonlocal flow for $U_4$. Hence the supersymmetrization procedure described here doesn't work for the $n = 4$ generalized KdV hierarchy.

It is a straightforward computation to work out the first hamiltonian structure for the $n$-KdV hierarchy and to show that the only cases in which the image of $J_1$ consists of perfect derivatives are indeed the $n = 2$ and $n = 3$ cases. Although this doesn't allow one to conclude that such supersymmetric extensions do not exist for $n > 3$, one can very easily compute the first nontrivial flow for the variable $u_4$. This calculation shows that $\frac{\partial u_4}{\partial t_2}$ will always contain two terms $(u_2')^2$ and $u_3 u_2'$ whose coefficients never vanish for $n > 3$ and which cannot be rewritten as perfect derivatives. Therefore, we conclude that the supersymmetrization of $n$-KdV induced by $u_i \mapsto U_i'$ does not work except for $n = 2$ and $n = 3$.

As a closing comment, let us remark that if, as expected, the SBs hierarchy has something to do with a supersymmetric analog of the two-matrix model, then we should expect some new feature to emerge in the generalization to the three-matrix model.

An Alternative Supersymmetrization of the $n$-KdV Hierarchy

As we saw in the previous section, what kills the supersymmetrization of the ($n$>3)-KdV hierarchy is the fact that the embedding $R[u_i] \to S[U_i]$ sends $u_i$ to $U_i'$, whence the need to strip off a derivative when deriving the equations for the superpartner of $u_i$. This entails the risk that the equations will not be local. For $n = 2$ and $n = 3$ this risk in not present, since the right hand side of the evolution equation (1) is always a perfect derivative; but for $n > 3$ this is not the case and one invariably obtains nonlocal equations.

The choice of embedding came dictated by the SKdV-B equation appearing in the matrix-model-inspired approach to two-dimensional quantum supergravity of [1] and therefore it would seem to be the choice to generalize when searching for supersymmetric integrable hierarchies that could possibly play a role in noncritical superstrings. From another point of view, however, this may not seem as natural a choice of embedding as $u_i \mapsto U_i$ where $U_i = u_i + \theta \sigma_i$ are primitive even superfields. It is clear that there is no longer any risk of obtaining nonlocal equations. We now summarize—without proof— some results about these hierarchies—hereafter denoted $n$-SKdV$'$. The diligent reader will be able to provide the proofs herself.

The following three results are analogous to Lemma 10, Proposition 12, and Proposition 17.

LEMMA 27. *Let $h(u_i) \in R[u_i]$. Then, in $S[U_i]$,*

$$\frac{\delta}{\delta U_i} \int_B h(U_i) = \frac{\delta}{\delta u_i} \int h(u_i) \bigg|_{u_i = U_i} .$$

PROPOSITION 28. *If $H_r \equiv \frac{n}{r} \operatorname{Tr} L^{r/n}$, $k \in \mathbf{N}$, are the conserved charges of the $n$-KdV hierarchy, then the corresponding conserved charges for its supersymmetrization are given by $K_r = \delta H_r$. They are nontrivial and independent for $r \not\equiv 0 \bmod n$.*

PROPOSITION 29. *The $n$-SKdV$'$ hierarchy inherits a bihamiltonian structure from that of $n$-KdV hierarchy.*

And as a corollary of these results, we have

COROLLARY 30. *The $n$-SKdV$'$ hierarchy is (formally) integrable.*

## ACKNOWLEDGEMENTS

We are grateful to E. Ramos for a careful reading of the manuscript. In addition, S.S. is grateful to W. Nahm and V. Rittenberg for their kind support and encouragement.